\documentclass[showpacs,amsmath,amssymb,aps,showkeys,floatfix,prd,a4paper]{revtex4}
\usepackage{graphicx} 
\usepackage{dcolumn}
\usepackage{bm}
\usepackage{epsfig}
\usepackage{amsfonts}
\usepackage{amssymb,amscd}
\usepackage{subfigure}
\usepackage{xcolor}

\def\lsim{\raise0.3ex\hbox{$<$\kern-0.75em\raise-1.1ex\hbox{$\sim$}}}

\def\gsim{\raise0.3ex\hbox{$>$\kern-0.75em\raise-1.1ex\hbox{$\sim$}}}

\newcommand{\be}{\begin{equation}}

\newcommand{\ee}{\end{equation}}

\def\beq{\begin{equation}}

\def\eeq{\end{equation}}

\def\beqa{\begin{eqnarray}}

\def\eeqa{\end{eqnarray}}

\newcommand{\ba}{\begin{eqnarray}}

\newcommand{\rr}{\mbox{\boldmath $r$}}

\newcommand{\rb}{\mbox{\boldmath $b$}}

\def\gappeq{\mathrel{\rlap {\raise.5ex\hbox{$>$}}

{\lower.5ex\hbox{$\sim$}}}}

\def\lappeq{\mathrel{\rlap{\raise.5ex\hbox{$<$}}

{\lower.5ex\hbox{$\sim$}}}}

\def\Toprel#1\over#2{\mathrel{\mathop{#2}\limits^{#1}}}

\begin{document}


\title{Leading neutron production at the EIC and LHeC: \\ estimating the impact of the absorptive corrections }


\author{F. Carvalho$^{1}$, V.P. Gon\c{c}alves$^{2}$,
  F.S. Navarra$^{3}$ and D. Spiering$^{3}$}


\affiliation{$^1$Departamento de Ci\^encias Exatas e
  da Terra, Universidade Federal de S\~ao Paulo,\\  
  Campus Diadema, Rua Prof. Artur Riedel, 275, Jd. Eldorado,
  09972-270, Diadema, SP, Brazil.\\ 
  $^2$ High and Medium Energy Group, Instituto de F\'{\i}sica e Matem\'atica,
  Universidade Federal de Pelotas\\
Caixa Postal 354,  96010-900, Pelotas, RS, Brazil.\\
$^3$Instituto de F\'{\i}sica, Universidade de S\~{a}o Paulo,
C.P. 66318,  05315-970 S\~{a}o Paulo, SP, Brazil.\\
}

\begin{abstract}
  Leading neutron (LN) production in $ep$ collisions at high energies is      
  investigated using the color dipole formalism and taking into account
  saturation effects. We update the  treatment of  absorptive effects  
  and estimate the impact of these effects on LN spectra
  in the kinematical range that will be probed by  the Electron Ion   
  Collider (EIC) and by the Large Hadron electron Collider (LHeC). We
  demonstrate that  Feynman scaling, associated to saturation, is not 
  violated  by the inclusion of absorptive effects. Moreover, our
  results indicate that the LN spectrum is strongly      
  suppressed at small photon virtualities. These results suggest that  
  absorptive  effects cannot be disregarded in future measurements of
  the $\gamma \pi$ cross section to be extracted from data on  leading
  neutron production.
\end{abstract}




\maketitle

\vspace{0.5cm}



\begin{figure}[t]
\begin{tabular}{ccc}
\includegraphics[width=.45\linewidth]{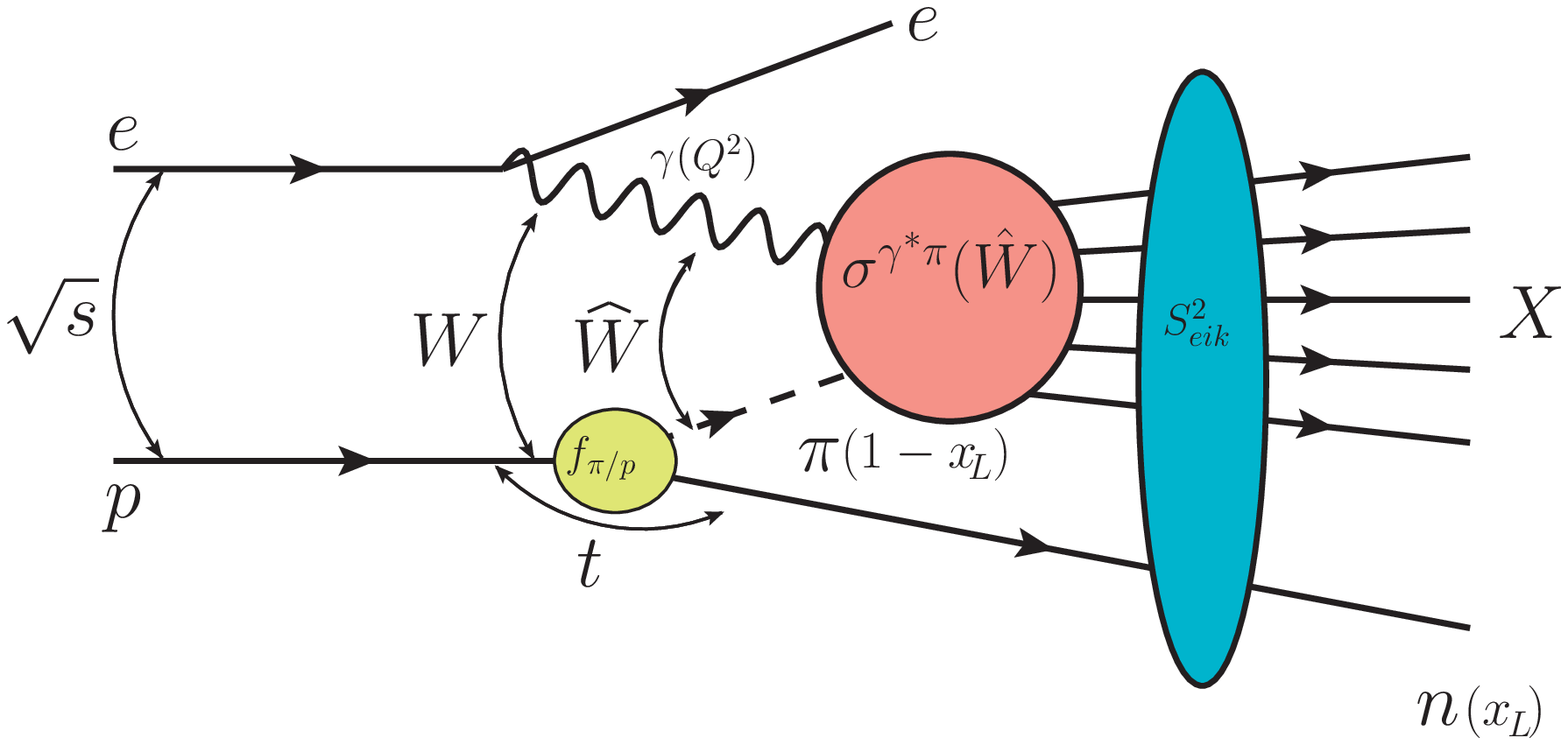}& \,\,\,\,\,\,&
  \includegraphics[width=.45\linewidth]{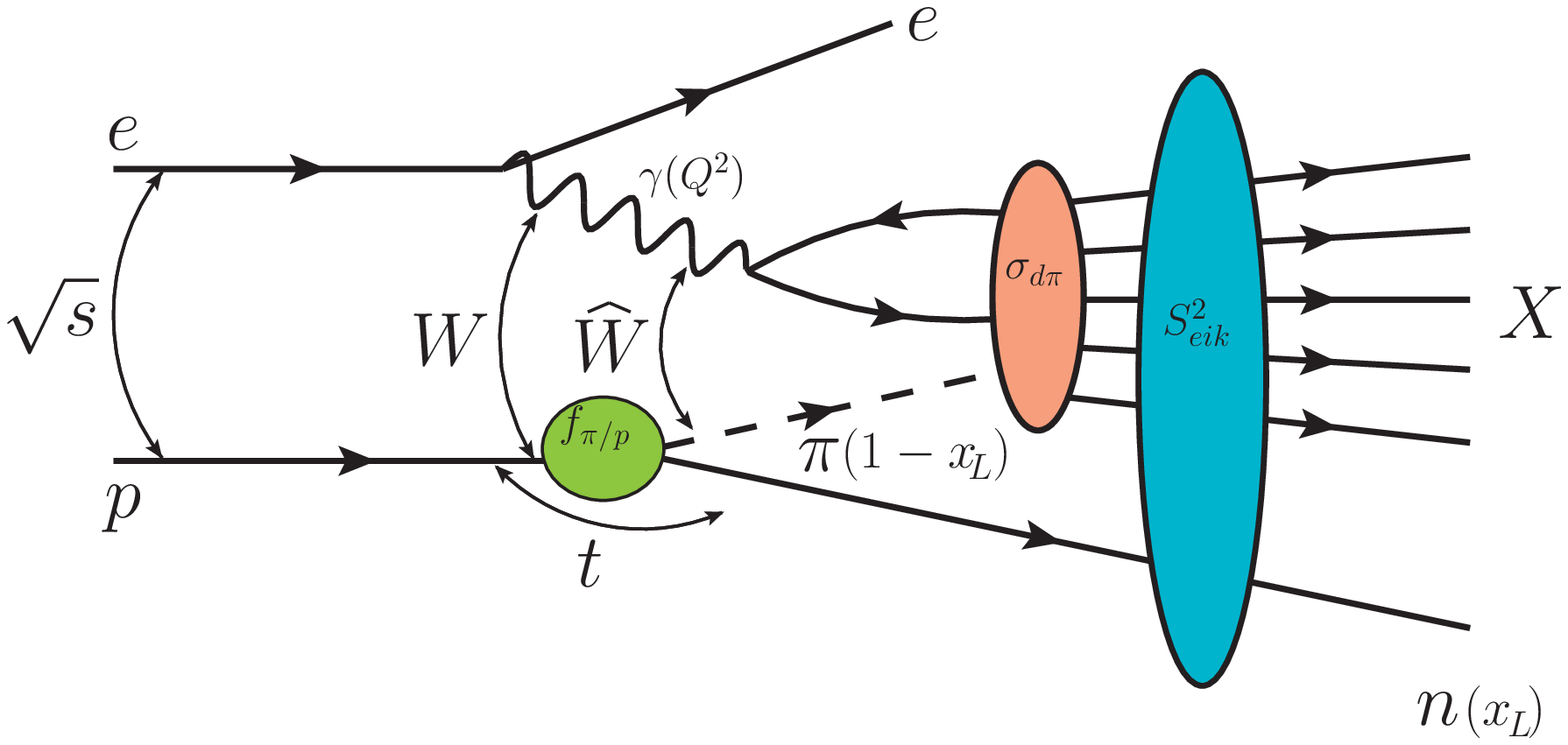} \\
  (a) & \,\,\,\,\,\, & (b)
	\end{tabular}
\caption{ (a) Leading neutron $n$ production in  $e p \rightarrow e n X $
  interactions at high energies. (b) Description of the process in the color
  dipole model.}
\label{Fig:diagram}
\end{figure}

The study of  deep inelastic  electron - proton ($ep$)  scattering has significantly improved our understanding of the proton structure in the high energy (small - $x$) regime (For a recent review see, e.g. Ref. \cite{rmp}).
In the future the partonic structure of other hadrons  will be
investigated \cite{pionkaon}. The pion structure has been discussed by several authors \cite{Holtmann,Kopeliovich:1996iw,Przybycien:1996z,Nikolaev:1997cn,holt,Kopeliovich:2012fd,McKenney:2015xis} and the subject became recently a hot topic due to the perspective of measuring the pion structure function $F_2^{\pi}(x,Q^2) $ in  future electron - hadron colliders at the BNL and CERN \cite{eic,lhec}.  The basic idea is that the pion structure can be probed in electron - proton collisions through the Sullivan process \cite{Sullivan:1971kd}, where the electron scatters off the meson cloud of the proton target. The associated processes  can be separated by tagging a forward neutron in the final state, which carries a large fraction of the proton energy.
Theoretically, this leading neutron production, is usually described assuming that the splitting  $p \rightarrow \pi^+ n$  and  the photon -- pion
interaction  can be factorized, as represented in Fig. 
\ref{Fig:diagram} (a), where $f_{\pi/p}$ represents the pion flux. Assuming the validity of the factorization hypothesis and the universality of the fragmentation process, which allows us to constrain $f_{\pi/p}$ using the data of leading neutron production in $pp$ collisions, we can obtain $\sigma^{\gamma^* \pi}$ and, consequently,  determine the $x$ and $Q^2$ dependencies of the pion structure function. However, the validity of this procedure is limited  by  absorptive effects, denoted by $S^2_{eik}$ in Fig. \ref{Fig:diagram},  that are associated to soft rescatterings between the produced and spectator particles. The studies performed in Refs. \cite{pirner,kop,Khoze:2017bgh}   indicated that these effects strongly affect leading neutron production in $pp$ collisions. In contrast, the absorptive corrections are predicted to be smaller in $ep$ collisions and their effects become weaker at larger photon virtualities \cite{Nikolaev:1997cn,pirner,Kaidalov:2006cw,Khoze:2006hw,Kopeliovich:2012fd,levin}. Although the treatment  of the absorptive corrections has advanced in recent years, they are still one of the main uncertainties in the study of  leading neutron production in  $pp$ collisions at RHIC and LHC and $ep$ collisions at the EIC and LHeC.

In Refs. \cite{nos1,nos2} we proposed a model to treat  leading neutron production in  $ep$ processes based on the color dipole formalism \cite{nik}. In this model, the virtual photon - pion cross section can be factorized in terms of the photon wave function (which describes the photon splitting in a $q\bar{q}$ pair) and the dipole - pion cross section $\sigma_{d\pi}$, as represented in Fig. \ref{Fig:diagram} (b). As shown in Refs. \cite{nos1,nos2}, the HERA data are quite well described by this approach assuming that absorptive corrections can be factorized  and represented by a multiplicative constant factor, denoted by ${{K}}$ in Ref. \cite{nos1}. Although successful
(in the limited kinematical range probed by HERA) and a
reasonable assumption to obtain a first approximation of the cross sections for the EIC and LHeC, it is fundamental to improve the description of $S^2_{eik}$  in order to derive more realistic predictions. Our goal in this paper is to revisit and update the approach proposed in Ref. \cite{pirner} for the absorptive effects. This approach allows us to estimate  these effects in terms of the color dipole formalism, i.e. using the same ingredients of the model proposed in  \cite{nos1,nos2}. As a consequence, we will be able to  derive parameter free predictions for the cross sections, which can be directly compared with the HERA data. Moreover, we will estimate the strength of the absorptive effects for different photon virtualities and center - of - mass energies and present predictions for  leading neutron production in future colliders.

Initially, let us discuss the approach proposed in Ref. \cite{nos1} to treat the  leading neutron production in $ep$ collisions, disregarding the absorptive effects.  At
high center - of - mass energies, this process  can be seen 
as a set of three factorizable subprocesses [See Fig. \ref{Fig:diagram} (b)]:  i) the photon emitted by the electron fluctuates into a 
quark-antiquark pair (the color dipole), ii) the color dipole interacts with the pion  and iii) the leading neutron is formed. In the color dipole formalism, the differential cross section reads: 
\begin{eqnarray}
  \frac{d^2 \sigma(W,Q^2,x_L,t)}{d x_L d t} & = & f_{\pi/p} (x_L,t) \,
  \sigma_{\gamma^* \pi}(\hat{W}^2,Q^2) \,\,, \\
& = & f_{\pi/p} (x_L,t) \times \int _0 ^1 dz \int d^2 \rr \sum_{L,T} \left|\Psi_{T,L} (z, \rr, Q^2)\right|^2  \sigma_{d\pi}({x}_{\pi}, \rr)
\label{crossgen}
\end{eqnarray}
where $Q^2$ is the virtuality of the exchanged photon,
$x_L$ is the proton momentum fraction carried by the 
neutron and $t$ is the square of the four-momentum of the exchanged pion. Moreover, 
 $\hat{W}$ is the center-of-mass energy of the 
virtual photon-pion system, which can be written as $\hat{W}^2 = (1-x_L) \, W^2$, where $W$ is the center-of-mass energy of the 
virtual photon-proton system.  In terms of the measured quantities $x_L$ and 
transverse momentum $p_T$ of the neutron, the pion 
virtuality is:
\beq
t \simeq-\frac{p_T^2}{x_L}-\frac{(1-x_L)(m_n^2-m_p^2 x_L)}{x_L} \,\,.
\label{virtuality}
\eeq
In Eq. (\ref{crossgen}), the virtual photon - pion cross section was expressed in terms  of the transverse and longitudinal photon wave functions $\Psi_i$, which
describe the photon splitting into  a $q\bar{q}$ pair of size $r \equiv |\rr|$,  and the dipole-pion cross section  $\sigma_{d\pi}$, which is determined by the QCD dynamics at high energies \cite{hdqcd}. 
The variable $z$ represents  the  longitudinal photon momentum fraction carried 
by the quark, the variable $\rr$ defines the relative transverse separation of the pair (dipole) and the scaling variable $x_{\pi}$ is defined by $x_{\pi} = x / (1-x_L)$, where $x$ is the Bjorken variable.

The flux factor $f_{\pi/p}$  gives the probability of the
splitting of a proton into a pion-neutron system and can be expressed as follows (See e.g. Ref. \cite{pirner}) 
\beq
f_{\pi/p}(x_L,t) = \frac{2}{3}\pi\sum_{\lambda\lambda'}
        |\phi_{n\pi}^{\lambda\lambda'}(x_L,{\bf p}_T)|^2
\eeq
where $\phi_{n\pi}^{\lambda\lambda'}(x_L,{\bf p}_T)$ is the probability
amplitude to find, inside a proton with spin up, a neutron with longitudinal
momentum fraction $x_L$, transverse momentum ${\bf p}_T$ and helicity 
$\lambda$ and a pion, with longitudinal momentum fraction $1-x_L$, 
transverse momentum $-{\bf p}_T$ and helicity $\lambda'$. In the light-cone approach, the amplitudes $\phi_{n\pi}$ of a proton with spin $+1/2$, read:
\beqa
\label{phi}
\phi_{n\pi}^{1/2,0}(x_L,{\bf p}_T) & = & 
          \frac{\sqrt{3}g_0}{4\pi\sqrt{\pi}}\frac{1}{\sqrt{x_L^2(1-x_L)}}
          \frac{m_n(x_L-1)}{M_{n\pi}^2-m_n^2}\nonumber\\
\phi_{n\pi}^{-1/2,0}(x_L,{\bf p}_T) & = & 
          \frac{\sqrt{3}g_0}{4\pi\sqrt{\pi}}\frac{1}{\sqrt{x_L^2(1-x_L)}}
          \frac{|{\bf p}_T|e^{-i\varphi}}{M_{n\pi}^2-m_n^2}\,,
\eeqa
where $M_{n\pi}^2$ is the invariant mass of the pion-neutron system,
given by
\[
M_{n\pi}^2 = \frac{m_n^2+p_T^2}{x_L} + \frac{m_\pi^2+p_T^2}{1-x_L}\,,
\]
with  $m_n$ and $m_\pi$ being the neutron and the pion masses, $g_0$ is the bare 
pion-nucleon coupling constant and $\varphi$ is the azimuthal angle
in the transverse plane. 
Because of the extended nature of the hadrons involved, the interaction
amplitudes in the above equations have to be
modified by including a phenomenological $\pi NN$ form factor, $G(x_L,p_T)$. It is
important to stress here that while the vertex is
derived from an effective meson-nucleon Lagrangian, the
form factor is introduced ad hoc. In our analysis we will choose the
covariant form factor, corrected by the Regge factor, given by 
\beq
\label{covff}
G(x_L,p_T)  =  {\rm exp}[R_{c}^2(t-m_\pi^2)] \, (1-x_L)^{-t}
\eeq
where $R_c^2 = 0.3$ GeV$^2$ was constrained using the HERA data (For details see Ref. \cite{nos1}).
The amplitude $\phi_{n\pi}^{\lambda\lambda'}(x_L,{\bf p}_T)$  changes to
$\phi_{n\pi}^{\lambda\lambda'}(x_L,{\bf p}_T) \, G(x_L,p_T)$ and then the
pion flux becomes: 
\beq
\label{flux}
f_{\pi/p}(x_L,t) = \frac{2}{3}\pi\sum_{\lambda\lambda'}
        |\phi_{n\pi}^{\lambda\lambda'}(x_L,{\bf p}_T)|^2|G(x_L,p_T)|^2\,,
\eeq
where $2/3$ is the isospin factor and the azimuthal angle in the transverse plane
has been integrated out.

In order to include the absorptive effects  in our predictions for the leading neutron spectrum $d\sigma/dx_L$, we will follow the approach proposed 
in Ref. \cite{pirner}, where these effects were estimated using the high - energy Glauber approximation \cite{glauber} to treat the multiple scatterings between the dipole and the pion -- neutron system. As demonstrated in Ref. \cite{pirner}, such approach can be easily implemented in the impact parameter space, implying that the spectrum can be expressed as follows
\beqa
\label{desdzgamma}
\frac{d\sigma(W,Q^2,x_L)}{dx_L} & = & 
        \int\! d^2{\rb}_{rel} \, \rho_{n\pi}(x_L,{\rb}_{rel})\,
        \int\! dz \, d^2{\rr} \,\sum_{L,T} \left|\Psi_{T,L} (z, \rr, Q^2)\right|^2  \sigma_{d\pi}({x}_{\pi}, \rr) \,
        S_{eik}^2(\rr,\rb_{rel}) \,\,\,,
\eeqa
where $ \rho_{n\pi}(x_L,{\rb}_{rel})$ is the probability density of
finding a neutron and a pion with momenta $x_L$ and $1-x_L$, respectively, and with a
relative transverse separation $\rb_{rel}$, which is given by  
\beq
\rho_{n\pi}(x_L,{\rb}_{rel}) = \sum_i|\psi^i_{n\pi}(x_L,{\rb}_{rel})|^2\,.
\label{rho}
\eeq
with 
\beq
\psi^i_{n\pi}(x_L,{\rb}_{rel}) = \frac{1}{2\pi}\int\!d^2{\bf p}_T \,  
        e^{i{\rb}_{rel}\cdot{\bf p}_T} \, \phi^i_{n\pi}(x_L,{\bf p}_T) \,,
\eeq
and $\phi^i_{n\pi}$ = $\sqrt{2/3} \, \phi^{\lambda\lambda'}_{n\pi} \, G(x_L,p_T)$.  Moreover, the survival factor $S_{eik}^2$ associated to the absorptive effects is expressed in terms of the  dipole -- neutron ($\sigma_{dn}$) cross sections as follows 
\begin{eqnarray}
S_{eik}^2(\rr,\rb_{rel}) = \Big\{1-\Lambda_{\rm eff}^2\frac{\sigma_{dn}(x_n,{\rr})}{2\pi}\,
    {\rm exp}\Big[-\frac{\Lambda_{\rm eff}^2{\rb}_{rel}^2}{2}\Big]\Big\}\,,
\label{Eq:esse2}    
\end{eqnarray}
where $x_n = x/x_L$ and $\Lambda^2_{\rm eff}$ is an effective parameter that was found to be equal to $0.1$ GeV$^2$ in Ref. \cite{pirner}. In our analysis, we will assume that $\sigma_{dn}$ is equal to the dipole - proton cross section, $\sigma_{dp}$, constrained by the HERA data. Finally, in order to estimate the spectrum, we must
specify the dipole - pion cross section, which is dependent on the description of the QCD dynamics at small - $x$. As in Ref. \cite{nos1}, we will assume that this quantity can be related to the dipole - proton cross section  using the additive quark model. Moreover, $\sigma_{dp}$ will be  described by the Color Glass Condensate (CGC) formalism, as given in the phenomenological model proposed in Ref. \cite{iim}. As a consequence, we will have that:
\begin{eqnarray}
\sigma_{d\pi} (x,\rr) =  \frac{2}{3} \cdot \sigma_{dp} ({x}, \rr) =  \frac{2}{3} \cdot  2 \pi R_p^2  \times  \left\{ \begin{array}{ll} 
{\mathcal N}_0\, \left(\frac{r\, Q_s}{2}\right)^{2\left(\gamma_s + 
\frac{\ln (2/r\, Q_s)}{K \,\lambda \, Y}\right)}\,, & \mbox{for $r 
Q_s({x}) \le 2$}\,,\\
 1-\text{e}^{-a\,\ln^2\,(b\,r\, Q_s)}\,,  & \mbox{for $r Q_s({x})  > 2$}\,, 
\end{array} \right.
\label{Eq:sigdp}
\end{eqnarray}
where  $a$ and $b$ are determined by continuity conditions at $r Q_s({x})=2$. The parameters  
$\gamma_s= 0.7376$, $\kappa= 9.9$, ${\mathcal N}_0=0.7$ and  $R_p = 3.344$ GeV$^{-1}$ has been adjusted using the HERA data in Ref. \cite{soyez}, with
the saturation scale $Q_s$  being given by: 
\beq
Q^2_s ({x}) = Q^2_0 \left( \frac{x_0}{{x}}\right)^{\lambda}
\label{qsat}
\eeq
with $x_0=1.632\times 10^{-5}$, $\lambda=0.2197$, $Q_0^2 = 1.0$ GeV$^2$.
The first line of Eq. (\ref{Eq:sigdp}) describes the linear regime whereas
the second one includes saturation effects.

\begin{figure}
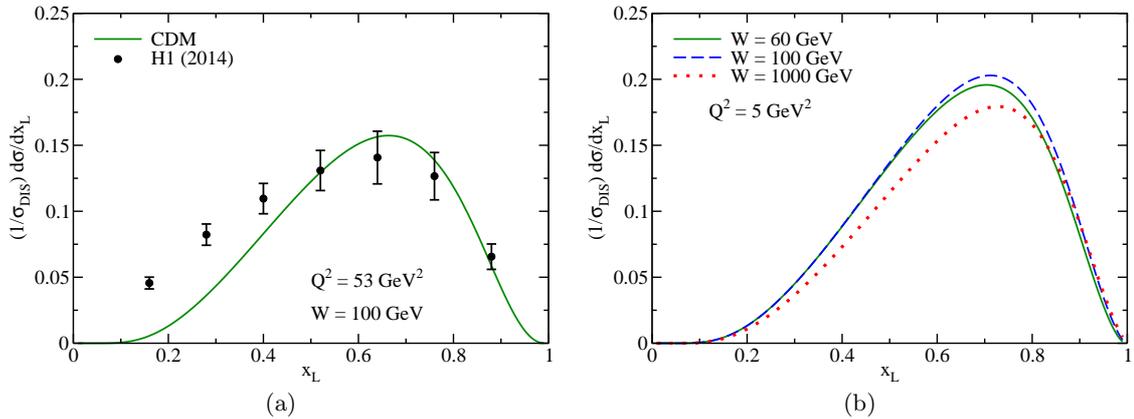

\begin{tabular}{ccc}
 \includegraphics[width=.40\linewidth]{abso_2a.eps}& \,\,\, &
  \includegraphics[width=.40\linewidth]{abso_2b.eps} \\
  (a) & \,\,\, & (b)
   \end{tabular}
\caption{(a) Comparison of the CDM prediction with the H1 data \cite{hera1}.
  (b) Predictions for the spectra considering different center - of - mass
  energies and $Q^2 = 5$ GeV$^2$.}
\label{Fig:comp}
\end{figure}

With the ingredients introduced above, we are ready to obtain parameter    
free predictions that can be compared with the HERA data. We can also derive
predictions which can be tested in  future $ep$ colliders.
In Fig. \ref{Fig:comp} (a) the CDM prediction for the kinematical range probed
by HERA is presented. As it can be seen, the H1 data \cite{hera1} are quite well 
described in the region $x_L \gtrsim 0.5$. As shown in previous studies
\cite{kop,Kaidalov:2006cw}, for smaller values of $x_L$, additional contributions
are expected to play a significant role. We can estimate the leading neutron 
spectrum for a kinematical range beyond that probed by HERA. We are particularly
interested in smaller values of the photon virtuality, where we expect a larger
contribution of       
saturation effects, and for the center - of - mass energies that will be 
reached at the EIC and LHeC. The results are presented in Fig. \ref{Fig:comp} (b). 
From the figure we see that the predictions are not strongly dependent on $W$. 
This is expected from the results presented in Ref. \cite{nos1}, where we have
demonstrated that  saturation leads to Feynman scaling, i.e. the energy 
independence of the $x_L$ spectra. Such scaling is expected to be strict when
the saturation scale becomes larger than the photon virtuality, which is 
satisfied for small values of $Q^2$ ($\lesssim 2$ GeV$^2$). However, as shown 
e.g. in Ref. \cite{iim}, the presence of the saturation effects also modifies
the behavior of the cross sections in a larger $Q^2$ range, implying the result
observed in Fig. \ref{Fig:comp} (b). In contrast, the DGLAP evolution leads to 
stronger violation of Feynman scaling, as shown in  Ref. \cite{nos1}. In 
a future experimental analysis of the leading neutron spectrum it will be very
interesting  to test this prediction of the Color Dipole Model.

\begin{figure}[t]
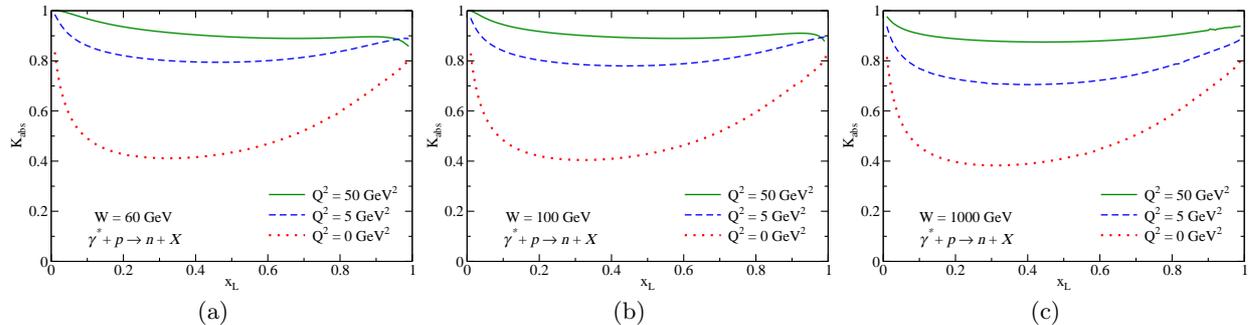

\begin{tabular}{ccc}
  \includegraphics[scale=0.22]{abso_3a.eps} &
{\includegraphics[scale=.22]{abso_3b.eps}} &
  \includegraphics[scale=0.22]{abso_3c.eps} \\
  (a) & (b) & (c)
   \end{tabular}
\caption{Dependence of the absorptive effects on $x_L$ in leading neutron
  production in $ep$ collisions for differents values of the photon virtuality
  and (a) $W = 60$ GeV, (b) $W = 100$ GeV and (c) $W = 1000$ GeV.}
\label{Fig:kfac}
\end{figure}

As discussed above, in order to measure the $\gamma \pi$ cross section and extract  
the pion structure function, it is crucial to have control of the absorptive effects
in the kinematical range probed by the collider. In particular, we should know the 
dependence of these effects on $Q^2$, $W$ and $x_L$. We can estimate the impact of
the absorptive effects through the calculation of the ratio between the cross
sections with and without absorption, where the latter is estimated assuming  
$S^2_{eik} = 1$. Our predictions for this ratio, denoted $K_{abs}$ hereafter,
are presented in Fig. \ref{Fig:kfac}. Our results show that the impact
increases for smaller values of $Q^2$ and larger energies $W$. For  $Q^2 = 50$
GeV$^2$, we see that  $K_{abs} \approx 0.9$ for $x_L \gtrsim 0.5$, with the
predictions being similar for the three values of $W$. This weak absorption is
expected in the Color Dipole Model, since at large values of $Q^2$ the main 
contribution for the cross section comes from dipoles with a small pair   
separation. In this regime, denoted color transparency, the impact of the
rescatterings is small, which implies that the absorptive effects become 
negligible. Another important aspect, is that for large photon virtualities, the
main effect of absorption is to suppress the cross section by a constant factor.
Similar results were derived in Ref. \cite{pirner}.   On the other hand, for
photoproduction ($Q^2 = 0$), we observe strong absorptive effects, which
reduce
the cross sections by a factor $\approx 0.4$ for $x_L = 0.5$. This result is
also expected, since for small $Q^2$ the cross section is dominated by large
dipoles and, consequently, the contribution of the rescatterings cannot be
disregarded. For larger values of $x_L$,  absorptive effects cannot be
modelled by a constant factor. Our conclusions agree with those derived in   
Ref. \cite{Kaidalov:2006cw} using Regge theory. Finally, our results     
indicate that the contribution of the absorptive effects is not strongly
energy dependent.
This result suggests that the main conclusion of Ref. \cite{nos1}, that the
spectra will satisfy  Feynman scaling, is still valid when the
absorptive effects are estimated using a more realistic model,
as already observed in Fig. \ref{Fig:comp} (b).

As a summary, in this paper we have updated the treatment of the absorptive
effects and incorporated them in the model proposed in our previous studies
\cite{nos1,nos2,nos3,nos4}, which is based on the color dipole formalism.
Using the approach proposed in Ref. \cite{pirner}, we have been able to
derive parameter free predictions for the leading neutron spectra. We
demonstrated that our model describes the HERA data in the region where
the pion exchange is expected to dominate. Moreover, we have presented
predictions for the kinematical ranges that will be probed by the future
EIC and LHeC. Our results indicate that the leading neutron spectra are not
strongly energy dependent at small photon virtualities. As shown in
Ref. \cite{nos1}, this almost energy independence (Feynman scaling) is a
consequence of saturation effects, which are expected to become significant  
at small - $Q^2$ and large energies. We have estimated the impact of the
absorptive effects, demonstrated that they increase at smaller photon
virtualities and that they depend on the longitudinal momentum $x_L$. 
Our  results show that  modelling  these effects by a constant factor is a
good approximation only for large $Q^2$. Our main conclusion is that a 
realistic measurement of the $\gamma \pi$ cross section in future colliders
and the extraction of the pion structure function  must take into account the
important contribution of the absorptive effects. Future experimental data on
leading neutron production in $ep$ collisions at the EIC              
will be crucial to test the main assumptions of our model, as well as
to improve our understanding of this important observable.

\begin{acknowledgments}
This work was  partially financed by the Brazilian funding
agencies CNPq, FAPESP,  FAPERGS and INCT-FNA (process number 
464898/2014-5).
 \end{acknowledgments}

\hspace{1.0cm}

\end{document}